\documentclass{article}

\usepackage{amssymb}
\usepackage{hyperref}
\usepackage{geometry}
\usepackage{amsfonts}
\usepackage{amsmath}
\usepackage{graphicx}

\begin{document}

\title{Non-Relativistic and Relativistic Equations for the Kratzer Potential
plus a Dipole in $2D$ Systems}
\author{{M. Heddar$^{1}$} \\
mebarek.heddar@univ-biskra.dz \and {\ M. Moumni$^{1,2}$} \\
m.moumni@univ-biskra.dz \and {\ M. Falek$^{1}$} \\
mokhtar.Falek@univ-biskra.dz\\
$^{1}$ Laboratory of Photonics and Nanomaterials (LPPNMM)\\
Department of Matter Sciences, University of Biskra, Algeria\\
$^{2}$ Laboratory of Radiations and Matter (PRIMALAB)\\
Department of Physics, University of Batna, Algeria}
\maketitle

\begin{abstract}
In this work, we study the wave equations in 2D Euclidian space for a new
non-central potential consisting of a Kratzer term and a dipole term $%
V\left( r,\theta\right) =Qr^{-1}+D_{r}r^{-2}+D_{\theta}\cos(\theta)r^{-2} $.
For Schrodinger equation, we obtain the analytical expressions of the
energies and the wave functions of the system. For Klein-Gordon and Dirac
equations, we do the study in both spin and pseudo-spin symmetries to get
the eigenfunctions and the eigenvalues. We also study the dependence of
energies on the parameters $D_{r}$ and $D_{\theta}$. We find that the $%
D_{\theta}$ term tends to dissociate the system, and thus counteracts the
Coulomb binding effect, and that the $D_{r}$ term can either amplify or
decrease this effect according to its sign.
\end{abstract}

\tableofcontents

\section{Introduction}

\label{Intro}

The quantum wave equations, namely Schr\"{o}dinger, Klein-Gordon and Dirac,
are very good toots for the study of atomic and molecular systems.
Unfortunately those equations have analytical solutions for very few type of
potentials and the majority are central. The complexity of the study of
multi-particle systems has forced theorists to use mean-field models and
most of them are also central. These central potentials don't correspond to
the reality of physical systems, except for the Coulomb potential that
matches perfectly the Hydrogen because it is just a two-particle system.

When we study non central potentials, the difficulty already appears in Schr%
\"{o}dinger equation, where the separation of variables is only possible
when the potential is of the type $V\left( r\right) +V\left( \theta\right)
r^{-2}+V\left( \varphi\right) r^{-2}\sin^{-2}\left( \theta\right) $ \cite%
{Khare94, Arda12, AlHaidari18, Kumari18}, and among them there is only few
ones that can be studied analytically \cite{Hautot73, Kumari18}. The study
of this kind of potential began with the works of Makarov \cite{Makarov67}
and Hartmann \cite{Hartmann72} in molecular systems and now they are in
focus in various fields, especially in quantum chemistry and nuclear
physics. For example they were used for the description of ring-shaped
molecules like benzene and also for the interactions between deformed pair
of nuclei \cite{Gharbi13, Bharali13, Sun14, Gribakin15}.

The most simple non-central potential is the electric dipole $D\cos\left(
\theta\right) r^{-2}$ (or pure dipole) and it attracted attention very early
because there were experimental clues that it could have bound states only
if its moment exceed a critical value \cite{Fermi47}. Experiments studies
followed and confirmed these results despite the fact that there is no
analytical solutions for this system \cite{Turner77}. This potential has
also attracted attention in chemistry given its applications in the anionic
bounds of polar molecules \cite{Fox66, Crawford67, Gutsev95, Jordan03} and
also in molecular biology in the study of electron binding mechanisms in DNA
and RNA \cite{Svozil04} (for a review see \cite{Simons08}).

From a theoretical point of view, the pure dipole potential has been studied
in the case of 3D systems \cite{Connoly07} and also in 1D and 2D systems
\cite{Connoly07, Glasser07}. Theoretical studies have also focused on the
potential of the non-pure dipole $Qr^{-1}+D\cos\left( \theta\right) r^{-2}$,
in 3D systems \cite{AlHaidari08} and in 2D ones \cite{Moumni16}. For this
last potential, both cases showed that the dipole must be below a critical
value to have bound states in the system; These results are in the opposite
of those of the pure dipole.

The interest for 2D systems comes from the great popularity of graphene (and
co. like silicene and manganene) and also from experimental achievements
with the realization of quantum gases at low dimensions \cite{Gorlitz01,
Martiyanov10} and before that from quasi-condensate experiments \cite%
{Safonov98}. In 2D systems, pure dipole is present in ultrathin
semiconductor layers \cite{Zhou90}, in spin-polarized atomic hydrogen
absorbed on the surface of superfluid helium \cite{Vasilyev02}, for charged
particles in a plane with perpendicular magnetic field \cite{Gadella11} and
also in gapped graphene with two charged impurities \cite{Martino14,
Klopfer14}. On the other hand, non-pure dipole potential was recently found
in the case of electron pairing that stems from the spin-orbit interaction
in two-dimensional quantum well \cite{Gindikin18}.

In this work we consider a new type of non-central potential consisting of a
combination of a Kratzer potential and a dipole term $V\left( r,\theta
\right) =Qr^{-1}+D_{r}r^{-2}+D_{\theta}\cos\left( \theta\right) r^{-2}$. Our
choice for this potential comes the fact that it is a better generalization
of the Coulomb interaction than the non-pure dipole potential considered in
\cite{AlHaidari08}, \cite{Moumni16} and \cite{Gindikin18} and also for the
potential richness of its applications given the number of parameters in its
expression. We add a dipole term to the Kratzer potential to take into
account the anisotropy of the distribution of charges and thus generalize
the studies already made for central distributions. Our study also aims to
add a new non-central system that has analytical solutions to those already
presented in the literature \cite{Hautot73, Kumari18}.

Kratzer potential was introduced to study the regularities in the band
spectra of diatomic molecules \cite{Kratzer20, Kratzer26}, and since then it
has attracted a lot of attention because of its applications in various
fields of physics and chemistry such as nuclear physics \cite{Fortunato03},
molecular physics \cite{Hagigeorgiou06}, quantum chemistry \cite{Berkdemir06}
and chemical physics \cite{Hooydonk09}. It is even used for the study of
optical properties in semiconductor quantum dots \cite{Batra18}. The studies
the Kratzer potential modified with angular terms followed for their
possible applications in ring-shaped organic molecules \cite{Cheng07,
Babaei11}. Recently, the Kratzer potential has been experimentally justified
in 2D systems because Rydberg series of s-type excitonic states in
monolayers of semiconducting transition metal dichalcogenides, which are 2D
semiconductors, follow a model system of 2D Kratzer type instead of a 2D
hydrogen atom \cite{Molas19}.

Our work will be structured as follows: after this introduction which
constitutes the first section \ref{Intro}, we will first start by a
theoretical justification of our choice of the non-central Kratzer potential
in the second section \ref{NCK}, then we will solve the Schr\"{o}dinger
equation for this new potential in the case of two-dimensional systems in
the third section \ref{2DNR}. Finally, after two sections devoted to the
results of the relativistic studies \ref{2DR}\ref{2DSR}, will come our
conclusions in the sixth section \ref{Concl}.

\section{The Non-Central Kratzer Potential}

\label{NCK}We consider a system composed of a point charge $q$\ in the
potential of a charge distribution $Q=\int\mathrm{d}q$\ (a cluster of point
charges $\mathrm{d}q$) such as an ion and a charged particle. The extended
charge $Q$ produce the following potential at the position of the test
charge $q$:%
\begin{equation}
V(\overrightarrow{r})=\int\frac{1}{4\pi\varepsilon_{0}}\frac{\mathrm{d}q_{a}%
}{r_{a}}  \label{1}
\end{equation}

We choose the origin $O$\ of the reference at the center of $Q$, We denote $%
M $ the position of $q$ and $\overrightarrow{r}$ the corresponding vector.
We put $\overrightarrow{A}$\ or $a$\ the position of the elementary charge $%
\mathrm{d}q$, so the relative position of the point charge $q$\ is $%
\overrightarrow{r}_{a}=\overrightarrow{AM}=\overrightarrow{OM}-%
\overrightarrow{OA}=\overrightarrow{r}-\overrightarrow{a}$; Thus we write:%
\begin{equation}
V(\overrightarrow{r})=\int\frac{1}{4\pi\varepsilon_{0}}\frac{\mathrm{d}q_{a}%
}{\left\Vert \overrightarrow{r}-\overrightarrow{a}\right\Vert }=\int\frac {1%
}{4\pi\varepsilon_{0}}\mathrm{d}q_{a}\left[ \left( \overrightarrow{r}-%
\overrightarrow{a}\right) ^{2}\right] ^{-1/2}  \label{2}
\end{equation}

We use Taylor series to write the integral because we consider that the
cluster's dimensions characterized by $\left\Vert \overrightarrow{a}%
\right\Vert $\ are small compared to the whole system represented by $%
\left\Vert \overrightarrow{r}\right\Vert $:%
\begin{equation}
V(\overrightarrow{r})=\int\frac{1}{4\pi\varepsilon_{0}}\frac{\mathrm{d}q_{a}%
}{r}\left[ \left( 1-2\frac{\overrightarrow{r}\cdot\overrightarrow{a}}{%
\overrightarrow{r}^{2}}+\frac{\overrightarrow{a}^{2}}{\overrightarrow{r}^{2}}%
\right) ^{2}\right] ^{-1/2}  \label{3}
\end{equation}

we restrict ourselves to the 1st order of the multipole expansion:%
\begin{equation}
\left[ \left( 1-2\frac{\overrightarrow{r}.\overrightarrow{a}}{%
\overrightarrow{r}^{2}}+\frac{\overrightarrow{a}^{2}}{\overrightarrow{r}^{2}}%
\right) ^{2}\right] ^{-1/2}=1+\frac{\overrightarrow{r}.\overrightarrow{a}}{%
r^{2}}+O\left( \frac{\overrightarrow{a}^{2}}{\overrightarrow{r}^{2}}\right)
\label{4}
\end{equation}

and we get:%
\begin{gather}
V(\overrightarrow{r})=\frac{1}{4\pi\varepsilon_{0}}\left( \int\frac {\mathrm{%
d}q_{a}}{r}+\int\frac{\mathrm{d}q_{a}}{r}\frac{\overrightarrow{r}.%
\overrightarrow{a}}{r^{2}}\right)  \notag \\
V(\overrightarrow{r})=\frac{1}{4\pi\varepsilon_{0}}\left( \frac{1}{r}\int%
\mathrm{d}q_{a}+\frac{1}{r^{2}}\int a\cos\theta_{a}\mathrm{d}q_{a}\right)
\label{5}
\end{gather}

The volume element in the integral is the one around the position $%
\overrightarrow{a}$\ of $\mathrm{d}q_{a}$, so the position $\overrightarrow{r%
}$ of the point charge $q$\ is a constant according to this integration and
the potential becomes:%
\begin{equation}
V(r)=\frac{1}{4\pi\varepsilon_{0}}\frac{Q}{r}+\frac{1}{4\pi\varepsilon_{0}}%
\frac{D_{r}}{r^{2}}  \label{6}
\end{equation}

$D_{r}$ represents the dipole moment of the cluster and it's denoted with
the indice $r$\ because one can consider it as a "radial" one or a dipole
with moment always pointing to the test charge $q$\ \cite{Sivoukhine86}. The
relation \ref{6} is equivalent to the Kratzer expression defined by:%
\begin{equation}
V_{K}(r)=d_{e}\left( \frac{r_{e}^{2}}{r^{2}}-\frac{2r_{e}}{r}\right)
\label{7}
\end{equation}

where we consider that $d_{e}$ is the dissociation energy and $r_{e}$ is the
equilibrium interatomic separation in the molecule \cite{Molski92, Durmus11,
Bao19}.

We see that the potential is central and this may not reflect reality
because the distribution is not usually perfectly symmetric. Therefore, we
have to take into account the possible anisotropy in the charge distribution
and to do this we consider that the positive and the negative centers of
charges do not coincide in $Q$ and we denote their positions $%
\overrightarrow{a}_{+} $\ and $\overrightarrow{a}_{-}$. This two centers
form an electric dipole representing this anisotropy and the potential of
such a dipole is just $D_{\theta}\cos(\theta)r^{-2}$. The dipole moment $%
D_{\theta}$ is proportional to the distance between the two charge centers
and the angle $\theta$\ defines the orientation of the position $%
\overrightarrow{r}$\ according to the dipole axis defined by $%
\overrightarrow{a}_{+}-\overrightarrow{a}_{-}$. We call this term the
"angular" dipole to differentiate it from the "radial" one.

Adding all the terms together gives us the Coulomb potential with two
dipoles and we call it a non-central (N-C) Kratzer potential:%
\begin{equation}
V\left( r,\theta\right) =\frac{1}{4\pi\varepsilon_{0}}\frac{Q}{r}+\frac {1}{%
4\pi\varepsilon_{0}}\frac{D_{r}}{r^{2}}+\frac{1}{4\pi\varepsilon_{0}}\frac{%
D_{\theta}\cos\theta}{r^{2}}  \label{8}
\end{equation}

This expression takes into account the non-point character of an ionic
system acting on an elementary charge and also the anisotropy in the charge
distribution of this ion; It represents the first correction of the Coulomb
effect of an extended charge.

\section{2D Schr\"{o}dinger Equation for N-C Kratzer Potential}

\label{2DNR}We write the two-dimensional stationary Schr\"{o}dinger equation
for a point charge in the potential \ref{8} and we use the polar coordinates
for the Laplacian:%
\begin{equation}
\left[ \frac{-\hbar^{2}}{2\mu}\left( \frac{\partial^{2}}{\partial r^{2}}+%
\frac{1}{r}\frac{\partial}{\partial r}+\frac{1}{r^{2}}\frac{\partial^{2}}{%
\partial\theta^{2}}\right) +\frac{q}{4\pi\varepsilon_{0}}\left( \frac {Q}{r}+%
\frac{D_{r}}{r^{2}}+\frac{D_{\theta}\cos\theta}{r^{2}}\right) \right]
\psi=E\psi  \label{9}
\end{equation}

We put the equation in the more convenient following form:%
\begin{equation}
\left[ \left( \frac{\partial^{2}}{\partial r^{2}}+\frac{1}{r}\frac{\partial
}{\partial r}-\frac{2\mu qQ}{4\pi\varepsilon_{0}\hbar^{2}}\frac{1}{r}-\frac{%
2\mu qD_{r}}{4\pi\varepsilon_{0}\hbar^{2}}\frac{1}{r^{2}}\right) +\frac{1}{%
r^{2}}\left( \frac{\partial^{2}}{\partial\theta^{2}}-\text{\ }\frac{2\mu
qD_{\theta}}{4\pi\varepsilon_{0}\hbar^{2}}\cos(\theta)\right) \right] \psi=%
\frac{-2\mu E}{\hbar^{2}}\psi  \label{10}
\end{equation}

and we write the solution as $\psi(r,\theta)=r^{-1/2}R(r)\Theta(\theta)$\ to
get two separate equations:
\begin{subequations}
\begin{gather}
\left( \frac{\partial^{2}}{\partial\theta^{2}}-E_{\theta}-\text{\ }\frac{%
2\mu qD_{\theta}}{4\pi\varepsilon_{0}\hbar^{2}}\cos(\theta)\right) \Theta
(\theta)=0  \label{11a} \\
\left[ \frac{\partial^{2}}{\partial r^{2}}+\left( E_{\theta}+\frac{1}{4}-%
\frac{2\mu qD_{r}}{4\pi\varepsilon_{0}\hbar^{2}}\right) \frac{1}{r^{2}}-%
\frac{2\mu qQ}{4\pi\varepsilon_{0}\hbar^{2}}\frac{1}{r}+\frac{2\mu E}{%
\hbar^{2}}\right] R(r)=0  \label{11b}
\end{gather}

We have to solve the angular equation \ref{11a} to find the constants $%
E_{\theta}$ and then we use these angular eigenvalues to solve the radial
equation \ref{11b}; this will give us the energies $E$\ of the system and
also the wave functions $\psi(r,\theta)$.

\subsection{Solution of Angular Equation}

The angular equation can easily be written as a Mathieu equation \cite%
{Mathieu} by defining $\theta=2z$, $a=-4E_{\theta}$ and $p=\mu
qD_{\theta}/\left( \pi\varepsilon_{0}\hbar^{2}\right) $:
\end{subequations}
\begin{equation}
\frac{\partial^{2}\Theta(z)}{\partial z^{2}}+(a-2p\cos2z)\Theta(z)=0
\label{12}
\end{equation}

The solutions of this equation are the cosine-elliptic $ce_{2m}\left(
z\right) $ and the sine-elliptic $se_{2m+2}\left( z\right) $\ functions
where $m$ is a natural number \cite{Abram72}. The solutions of the Mathieu
equation are periodic because $z$\ has $\pi $\ as a period and this lead us
to consider the Floquet's theorem \cite{Floquet} or the Bloch's theorem \cite%
{Bloch28}. They stipulate that, for a given value of the parameter $p$, the
solution is periodic only for certain values of the other parameter $a$;
They are called characteristic values and denoted $a\left( 2m,p\right) $ or $%
a_{2m}\left( p\right) $ for the cosine solutions and $b\left( 2m,p\right) $
or $b_{2m}\left( p\right) $ for the sine ones.

There is no analytical expression for the Mathieu characteristic values $%
a_{2m}\left( p\right) $ and $b_{2m}\left( p\right) $, so they are usually
given either numerically or graphically. This doesn't preclude that we can
write approximate analytical expressions for small and large values of $p$\
\cite{NIST}. For small values of $p$, we can express $a$ and $b$ for $m>3$\
as ($l=4m^{2}-1$):%
\begin{equation}
a_{2m}=b_{2m}\approx 4m^{2}+\frac{1}{2l}\,p^{2}+\frac{20m^{2}+7}{%
32l^{3}\left( l-3\right) }\,p^{4}+\frac{36m^{4}+232m^{2}+29}{64l^{5}\left(
l-3\right) \left( l-8\right) }\,p^{6}+\mathcal{O}\left( p^{8}\right)
\label{13}
\end{equation}

The coefficients of the power series of $a_{2m}\left( p\right) $ and $%
b_{2m}\left( p\right) $ are the same until the terms in $p^{2m-2}$.

We have similar polynomials for $m\leq 3$\ but with different coefficients
for the $a$'s and the $b$'s. We note here that there is no sine solutions
for $m=0$ and so there is no $b(m=0$).

For large values of $p$, we get another polynomial ($k=2n+1$):%
\begin{equation}
a_{n}=b_{n+1}\approx -2p+2kp^{1/2}-\frac{1}{8}\left[ k^{2}+1\right] -\left[
k^{3}+3k\right] \frac{1}{2^{7}p^{1/2}}-\left[ 5k^{4}+34k^{2}+9\right] \,%
\frac{1}{2^{12}p}+\mathcal{O}\left( p^{-3/2}\right)  \label{14}
\end{equation}

From now on to simplify the writing, we will use the same symbol $%
c_{2m}\left( p\right) $ for both the characteristic values $a_{2m}\left(
p\right) $ and $b_{2m}\left( p\right) $.

Using the definitions $a=-4E_{\theta }$ and $p=\mu qD_{\theta }/\left( \pi
\varepsilon _{0}\hbar ^{2}\right) $ together with the relation of Mathieu
parameters enable us to write the angular eigenstates as a function of the
angular moment:%
\begin{equation}
E_{\theta }^{(2m)}=-\frac{1}{4}c_{2m}\left( \frac{\mu q}{\pi \varepsilon
_{0}\hbar ^{2}}D_{\theta }\right)  \label{15}
\end{equation}

From \ref{13}, we see that for small values of $D_{\theta}$ (or $p$), the
angular solution can be put in the form:%
\begin{equation}
E_{\theta}^{(2m)}=-m^{2}+P_{m}(D_{\theta})  \label{16}
\end{equation}

where $P_{m}(D_{\theta})$\ is a polynomial in terms of even power of $%
D_{\theta}$\ starting from $2$. This expression will be used to validate our
solutions in the limit $D_{\theta}\rightarrow0$.

\subsection{Solution of Radial Equation}

Using the asymptotic behavior of the solutions to solve the radial equation %
\ref{11b}, we start with the transformation $R(r)=r^{\lambda}e^{-\beta
r}f(r) $, so we get a new differential equation for $f(r)$:%
\begin{equation}
\left[ r\frac{d^{2}}{dr^{2}}+2(\lambda-\beta r)\frac{d}{dr}-2\left( \frac{%
\mu qQ}{4\pi\varepsilon_{0}\hbar^{2}}+\lambda\beta\right) \right] f(r)=0
\label{17}
\end{equation}

Because parameters\ $\beta$\ and $\lambda$\ are free ones, we chose them as
follows to simplify the equation:%
\begin{equation}
\beta^{2}=-\frac{2\mu E}{\hbar^{2}}\text{\ \& }\lambda(\lambda-1)+E_{\theta
}^{(m)}-\frac{2\mu qD_{r}}{4\pi\varepsilon_{0}\hbar^{2}}+\frac{1}{4}=0
\label{18}
\end{equation}

As $\psi\left( r,\theta\right) $ must be convergent, the accepted solutions
for these parameters that let $R(r)$\ non-singular at $r=0$ are:%
\begin{equation}
\beta=\sqrt{-\frac{2\mu E}{\hbar^{2}}}\text{ \& }\lambda=\frac{1}{2}+\sqrt{%
-E_{\theta}^{(m)}+\frac{2\mu qD_{r}}{4\pi\varepsilon_{0}\hbar^{2}}}
\label{19}
\end{equation}

We can reduce \ref{17}\ to a confluent hypergeometric type by defining a new
variable $z=2\beta r$:%
\begin{equation}
\left[ z\frac{d^{2}}{dz^{2}}+(2\lambda-z)\frac{d}{dz}-(\frac{\mu qQ}{%
4\pi\varepsilon_{0}\hbar^{2}}\frac{1}{\beta}+\lambda)\right] f(z)=0
\label{20}
\end{equation}

The solution here is just the confluent hypergeometric function:%
\begin{equation}
f\left( z\right) =N_{1}F_{1}\left( \lambda+\frac{\mu qQ}{4\pi
\varepsilon_{0}\hbar^{2}}\beta^{-1},2\lambda,2\beta r\right)  \label{21}
\end{equation}

and the wave function of the system follows:%
\begin{equation}
\psi\left( r,\theta\right) =Nr^{\lambda-\frac{1}{2}}e^{-\beta r}\Theta\left(
\theta\right) _{1}F_{1}\left( \lambda+\frac{\mu qQ}{4\pi\varepsilon_{0}%
\hbar^{2}}\beta^{-1},2\lambda,2\beta r\right)  \label{22}
\end{equation}

We compute the normalization constant $N$\ from the condition $\int
\left\vert \psi(r,\theta)\right\vert ^{2}rdrd\theta=1$ and we recall that
Mathieu functions are normalized by definition to $\pi$\ \cite{Abram72}:%
\begin{equation}
N=\frac{2^{\lambda}\beta^{\lambda+\frac{1}{2}}}{(2\lambda-1)!\pi}\left(
\frac{(n_{r}+2\lambda-1)!}{n_{r}!(n_{r}+2\lambda)!}\right) ^{1/2}  \label{23}
\end{equation}

Here we have used Laguerre polynomials of degree $n_{r}$\ \cite{Abram72}:%
\begin{equation}
L_{n_{r}}^{2\lambda-1}(2\beta r)=\frac{(n_{r}+2\lambda-1)!}{%
n_{r}!(2\lambda-1)!}\text{ }_{1}F_{1}\left( -n_{r},2\lambda,2\beta r\right)
\label{24}
\end{equation}

and the identity \cite{Abram72}:%
\begin{equation}
\int_{0}^{\infty}e^{-q}q^{k+1}\left[ L_{n}^{k}\left( q\right) \right] ^{2}dq=%
\frac{\left( n+k\right) !}{n!}\left( 2n+k+1\right)  \label{25}
\end{equation}
One can also use the Appell's double series $F_{2}$\ to get the same
normalized constant \cite{Saad03}.

From the asymptotic behavior of the confluent series ($r\rightarrow
\infty\Longrightarrow$ $_{1}F_{1}=0$) which leads to $\psi\rightarrow0$ for $%
r\rightarrow\infty$, we find the condition of quantization:%
\begin{equation}
\lambda+\frac{\mu qQ}{4\pi\varepsilon_{0}\hbar^{2}}\beta^{-1}=-n_{r}\text{ ,
}n_{r}=0,1,2,...  \label{26}
\end{equation}

and we use this condition with the relation \ref{19}\ to obtain the spectrum
of the discrete energy levels of our system:%
\begin{equation}
E_{n_{r},m}=-\left[ \left( \frac{4\pi\varepsilon_{0}\hbar^{2}}{2\mu qQ}\sqrt{%
\frac{\hbar^{2}}{2\mu}}\right) \left( n_{r}+\frac{1}{2}+\sqrt {%
-E_{\theta}^{(m)}+\frac{2\mu qD_{r}}{4\pi\varepsilon_{0}\hbar^{2}}}\right) %
\right] ^{-2}  \label{27}
\end{equation}

Starting from this expression, we can get the solutions of the usual 2D
Kratzer potential \cite{Oyewumi05, Agboola11} by taking the limit $D_{\theta
}\rightarrow0$, so $P_{m}(D_{\theta})\rightarrow0$ which lead to $E_{\theta
}^{(m)}=-m^{2}$ from \ref{16}:%
\begin{equation}
E_{n_{r},m}^{(Kratzer)}=-\left[ \left( \frac{4\pi\varepsilon_{0}\hbar^{2}}{%
2\mu qQ}\sqrt{\frac{\hbar^{2}}{2\mu}}\right) \left( n_{r}+\frac{1}{2}+\sqrt{%
m^{2}+\frac{2\mu qD_{r}}{4\pi\varepsilon_{0}\hbar^{2}}}\right) \right] ^{-2}
\label{28}
\end{equation}

If we add the limit $D_{r}\rightarrow0$, we find the 2D Coulomb solutions
\cite{Zaslow67, Parfitt02}:%
\begin{equation}
E_{n_{r},m}^{(Coulomb)}=-\left( \frac{4\pi\varepsilon_{0}\hbar^{2}}{2\mu qQ}%
\sqrt{\frac{\hbar^{2}}{2\mu}}\right) \left( n_{r}+\left\vert m\right\vert +%
\frac{1}{2}\right) ^{-2}  \label{29}
\end{equation}

Comparing with these results, we use the notations of the 2D hydrogen atom $%
n=n_{r}+\left\vert m\right\vert $ and we write the energy eigenvalues of the
system ($E_{\theta }^{(m)}$ is replaced from \ref{15}):%
\begin{equation}
E_{n,m}=-\left[ \left( \frac{4\pi \varepsilon _{0}\hbar ^{2}}{\mu qQ}\sqrt{%
\frac{\hbar ^{2}}{2\mu }}\right) \left( n-\left\vert m\right\vert +\frac{1}{2%
}+\sqrt{\frac{1}{4}c_{2m}\left( \frac{\mu q}{\pi \varepsilon _{0}\hbar ^{2}}%
D_{\theta }\right) +\frac{2\mu qD_{r}}{4\pi \varepsilon _{0}\hbar ^{2}}}%
\right) \right] ^{-2}  \label{30}
\end{equation}

For our numerical computations, we use the same considerations as those of
molecular systems. So we choose the extended charge as a positive ion and
the point charge is an electron, and we get two opposite charges equal in
magnitude $q=-Q=-e$. We use the Hartree atomic units where $\hbar =e=\mu
=4\pi \varepsilon _{0}=1$ and the energies become:%
\begin{equation}
E_{n,m}=-2\left( n-\left\vert m\right\vert +\frac{1}{2}+\sqrt{\frac{1}{4}%
c_{2m}\left( 4D_{\theta }\right) +2D_{r}}\right) ^{-2}  \label{31}
\end{equation}

We note in this relation of the energies, that the angular dipole removes
the degeneracy of the sine and cosine states for $m\neq 0$. This degeneracy
is restored when the angular moment vanishes since the two Mathieu's
characteristic parameters $a_{2m}$ and $b_{2m}$ have the same limit in this
case \ref{13}. The result restores those of the ordinary Kratzer potential
(or Coulomb potential) where the wave function of each level $E_{n,m}$ is a
linear combination of both sine and cosine states. For the $s$-states ($m=0$%
), we only find the cosine solutions because the sine solutions are absent
in this case.

\begin{figure}
\centering
\includegraphics[width=0.5\textwidth]{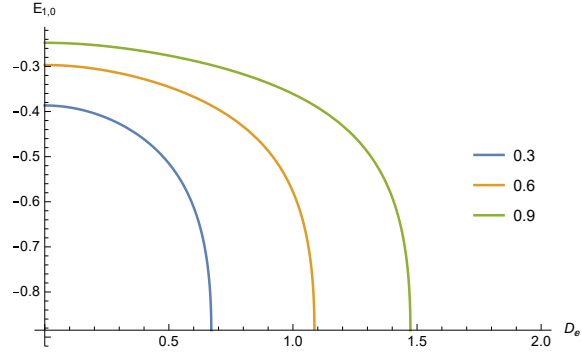}
\caption{$E(1,0)$ vs $D_\protect\theta$ for $D_r=0.3$,$0.6$ and $0.9$}
\label{fig1}
\end{figure}

\begin{figure}
\centering
\includegraphics[width=0.5\textwidth]{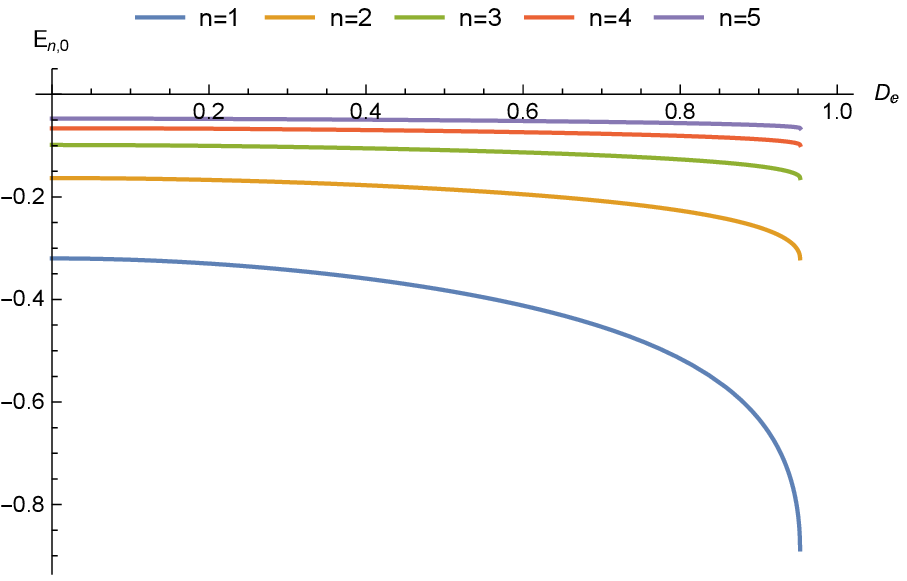}
\caption{$E(n,0)$ vs $D_\protect\theta$ for $D_r=0.5$ and $n=1$, $2$, $3$, $%
4 $ and $5$}
\label{fig2}
\end{figure}

\begin{figure}
\centering
\includegraphics[width=0.5\textwidth]{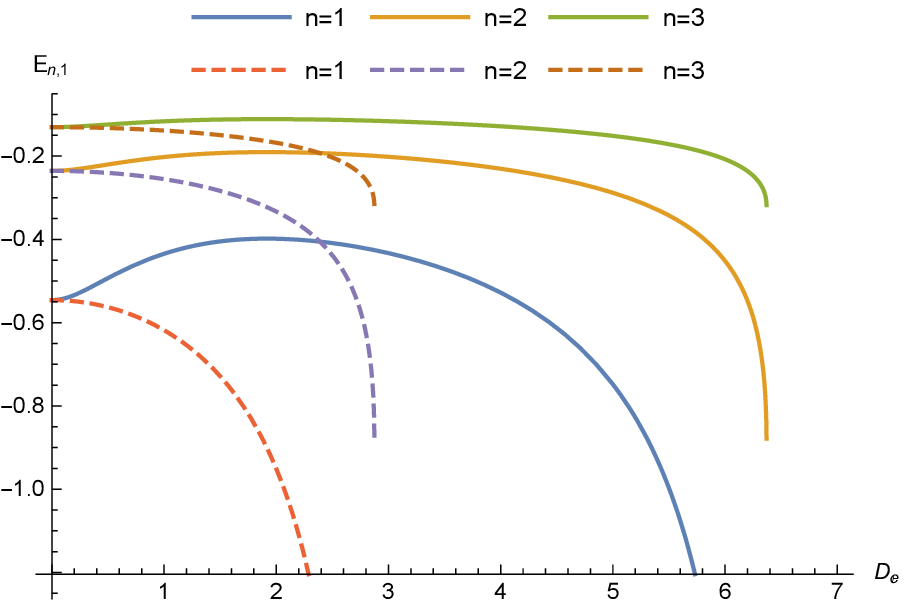}
\caption{$E(n,1)$ vs $D_\protect\theta$ for $D_r=0.5$ and $n=1$, $2$, $3$}
\label{fig3}
\end{figure}

\begin{figure}
\centering
\includegraphics[width=0.5\textwidth]{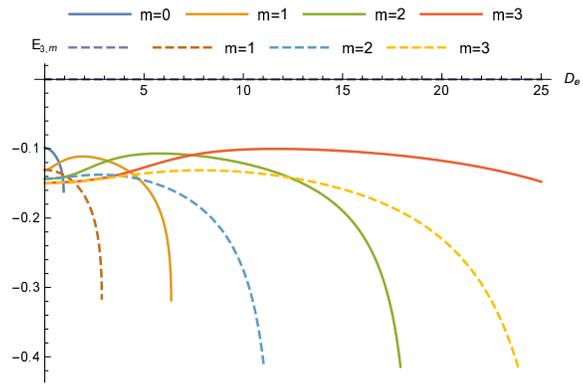}
\caption{$E(3,m)$ vs $D_\protect\theta$ for $D_r=0.5$ and $m=0$, $1$, $2$
and $3$}
\label{fig4}
\end{figure}

Through the expression \ref{31}, we see that the behavior of the energies
follows essentially that of the Mathieu's parameters and thus the angular
moment, whereas the effect of the radial moment merely shifts the energies
to larger or smaller values according to its sign. The sign of the angular
moment doesn't affect the results because the parameters $c_{2m}$\ are even
functions. Of course, the energies increase with the $n$ and decrease with
the $m$ but the main effect of the $m$ is to extend the allowed region for
the values of the angular momentum. We also note that the energies
corresponding to the $ce_{2m}\left( z\right) $ solutions are larger than the
$se_{2m}\left( z\right) $ ones and this is caused by the fact that the $%
a_{2m}$ are bigger than the $b_{2m}$. (see figures \ref{fig1}\ref{fig2}\ref{fig3}%
\ref{fig4} were dashed lines are for sine solutions).

The main remark that can be drawn from \ref{31} is that there is an
essential condition for the system to have bound states:%
\begin{equation}
\frac{1}{4}c_{2m}\left( 4D_{\theta }\right) +2D_{r}>0  \label{32}
\end{equation}

\begin{figure}
\centering
\includegraphics[width=0.5\textwidth]{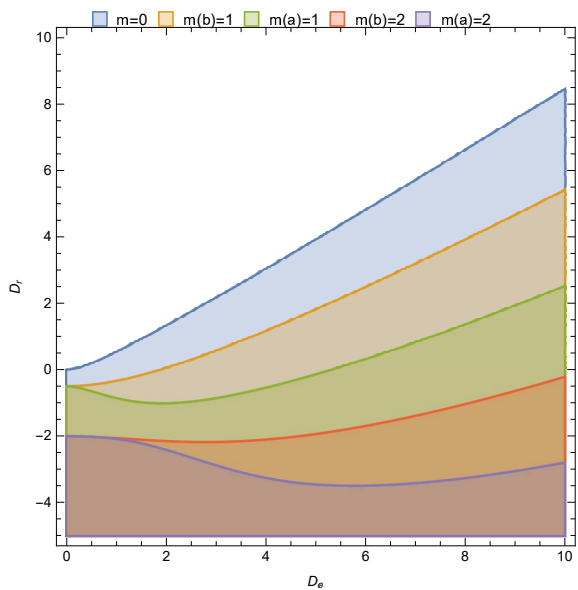}
\caption{Forbiden regions of $D_r$ and $D_\protect\theta$\ for $m=0,1,2$}
\label{fig5}
\end{figure}

\begin{figure}
\centering
\includegraphics[width=0.5\textwidth]{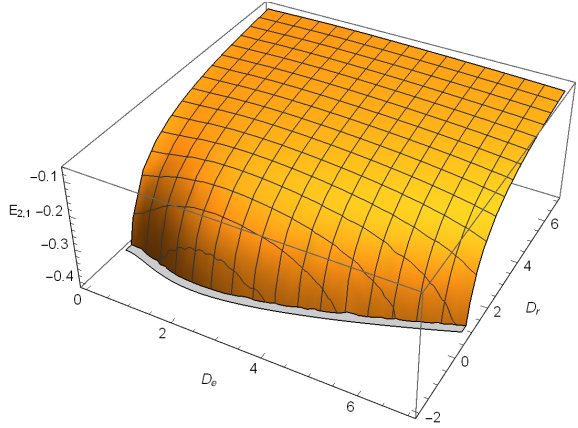}
\caption{$E(2,1)$ vs $D_r$ and $D_\protect\theta$}
\label{fig6}
\end{figure}

This condition shows that there are critical values for the two dipole
moments, depending only on the quantum number $m$, that make the
corresponding bound state no longer exists. If we put $D_{r}=0$, all the $s$%
-states ($m=0$) are absent because the critical value for $D_{\theta }$\
here is zero. We say here that the presence of radial dipole is essential
for $s$-states to exist, otherwise the angular moment make them disappear.
The same observation is made concerning the other $m$-states ($m>0$), but
the critical value of the angular moment is positive in all these cases and
these critical values increase with $m$\ and also with the values of $D_{r}$
(fig\ref{fig5}). This critical value is smaller for the sine states and this
causes the spread of the spectrum of these states to be less than that of
the cosine states on the axis of the angular momentum (figures \ref{fig4}
and \ref{fig5} where the indice a is for cosine solutions and the b for sine ones). So the radial dipole has two effects, it moves the energies
to higher values while enlarging the region of possible values of angular
moment (fig\ref{fig6}).

The table \ref{tab1} shows the critical values of $D_{\theta}$ for different
values of $D_{r}$ and $m$ when we consider cosine solutions.

\begin{table}
\centering
\begin{tabular}{ccccc}
$D_{r}\backslash m$ & $0$ & $1$ & $2$ & $3$ \\
$-0.3$ & $-$ & $1.925$ & $8.004$ & $17.462$ \\
$+0.0$ & $-$ & $2.662$ & $8.679$ & $18.132$ \\
$+0.3$ & $0.543$ & $3.284$ & $9.323$ & $18.782$ \\
$+0.6$ & $0.923$ & $3.851$ & $9.942$ & $19.420$ \\
$+0.9$ & $1.284$ & $4.385$ & $10.543$ & $20.046$%
\end{tabular}
\
\caption{Critical values for $D_{\protect\theta}$}
\label{tab1}
\end{table}

\section{2D Relativistic Equations for N-C Kratzer Potential}

\label{2DR}We start by considering the general case of the non-central
potential $V(r,\theta)=V\left( r\right) +V\left( \theta\right) /r^{2}$ in 2D
systems. There are no analytical solutions for the Klein-Gordon and Dirac
equations for this potentials, but both equations reduce to the same Schr%
\"{o}dinger type equation if we consider the cases of spin and pseudo-spin
symmetries. These symmetries have been considered for the first time in
shell models of nuclear physics, where protons and neutrons are treated in
the same way \cite{Hecht69, Arima69}. Then they became essential symmetries
in relativistic theories \cite{Ginocchio97, Zhou03} and since then their
studies have increased considerably in these systems \cite{Liang15}. These
symmetries were also used to study the relativistic theory of both central
and ring-shaped Kratzer potentials \cite{Hall10, Berkdemir08}.

We will consider the Klein-Gordon equation first then the Dirac equation and
we start with the spin symmetry case and then the pseudo-spin one.

\subsection{Klein-Gordon Equation}

The stationary Klein-Gordon equation for a single charge $q$ in both scalar $%
S\left( \vec{r}\right) $ and vector $U\left( \vec{r}\right) $ potentials is
written as:%
\begin{equation}
\left[ c^{2}p^{2}-\left( E-U\left( \vec{r}\right) \right) ^{2}+\left( \mu
c^{2}+S\left( \vec{r}\right) \right) ^{2}\right] \psi\left( \overrightarrow{r%
}\right) =0  \label{33}
\end{equation}
The denominations vector potential is used for the energy $U$ but it isn't
vectorial in any way. The vectorial potential $\overrightarrow{A}\left(
\overrightarrow{r}\right) $ comes with the momentum $\overrightarrow{p}$ and
is zero in our case.

Spin or pseudo-spin symmetry are defined by $S\left( \vec{r}\right) =\pm
U\left( \vec{r}\right) $ and the two reduce the wave equation \ref{33} to
the following second order equation:%
\begin{equation}
\left[ c^{2}p^{2}+2\left( E\pm\mu c^{2}\right) U\left( \vec{r}\right)
-\left( E^{2}-\mu^{2}c^{4}\right) \right] \psi\left( \overrightarrow{r}%
\right) =0  \label{34}
\end{equation}

The equation is easily written as a Schr\"{o}dinger equation with the
transformations:%
\begin{equation}
\left( \frac{E}{\mu c^{2}}\pm1\right) U\left( \vec{r}\right) \longrightarrow
U\left( \vec{r}\right) \text{ \& }\frac{1}{2}\left( \frac{E^{2}}{\mu c^{2}}%
-\mu c^{2}\right) \longrightarrow E  \label{35}
\end{equation}
Here we get a system where the potential depends on the energy. These energy
dependant potentials have been considered for a long time when the
relativistic effects began to be taken into account in quantum physics \cite%
{Pauli27, Snyder40, Shiff40}. Recently a lot of works were devoted to this
type of potentials \cite{Schulze17, Schulze18, Benzair18} (and the
references therein).

\subsection{Dirac Equation}

We consider now the stationary Dirac equation:%
\begin{equation}
\left[ c\overrightarrow{\alpha}\cdot\overrightarrow{p}+\beta\left( \mu
c^{2}+S\left( \vec{r}\right) \right) -\left( E-U\left( \vec{r}\right)
\right) \right] \psi\left( \overrightarrow{r}\right) =0  \label{36}
\end{equation}

and we use the Pauli-Dirac representation:%
\begin{equation}
\overrightarrow{p}=-i\hbar\overrightarrow{\nabla}\text{ };\text{ }%
\overrightarrow{\alpha}=\left(
\begin{array}{cc}
0 & \overrightarrow{\sigma} \\
\overrightarrow{\sigma} & 0%
\end{array}
\right) \text{ };\text{ }\beta=\left(
\begin{array}{cc}
I & 0 \\
0 & -I%
\end{array}
\right)  \label{37}
\end{equation}

where $\overrightarrow{\sigma}$ is the vector of Pauli matrices and $I$ is
the $2\times2$ identity matrix.

We write the wave function as a two component vector of the Pauli-Dirac
representation:%
\begin{equation}
\psi\left( \overrightarrow{r}\right) =\left(
\begin{array}{c}
\varphi\left( \overrightarrow{r}\right) \\
\chi\left( \overrightarrow{r}\right)%
\end{array}
\right)  \label{38}
\end{equation}

and we obtain two coupled differential equations:%
\begin{align}
c\overrightarrow{\alpha}\cdot\overrightarrow{p}\chi\left( \overrightarrow{r}%
\right) & =\left[ E-U\left( \vec{r}\right) -\mu c^{2}-S\left( \vec {r}%
\right) \right] \varphi\left( \overrightarrow{r}\right)  \notag \\
c\overrightarrow{\alpha}\cdot\overrightarrow{p}\varphi\left( \overrightarrow{%
r}\right) & =\left[ E-U\left( \vec{r}\right) +\mu c^{2}+S\left( \vec{r}%
\right) \right] \chi\left( \overrightarrow{r}\right)  \label{39}
\end{align}

If we consider spin symmetry, where $S\left( \vec{r}\right) =U\left( \vec{r}%
\right) $, the system is decoupled by writing:%
\begin{equation}
\chi\left( \overrightarrow{r}\right) =\frac{c\overrightarrow{\alpha}\cdot%
\overrightarrow{p}}{E+\mu c^{2}}\varphi\left( \overrightarrow{r}\right)
\label{40}
\end{equation}
and we get the following second order equation:%
\begin{equation}
\left[ c^{2}p^{2}+2\left( E+\mu c^{2}\right) U\left( \vec{r}\right) -\left(
E^{2}-\mu^{2}c^{4}\right) \right] \varphi\left( \overrightarrow{r}\right) =0
\label{41}
\end{equation}

In the same way, using pseudo-spin symmetry relation $S\left( \vec{r}\right)
=-U\left( \vec{r}\right) $, we write:%
\begin{equation}
\varphi\left( \overrightarrow{r}\right) =\frac{c\overrightarrow{\alpha}\cdot%
\overrightarrow{p}}{E-\mu c^{2}}\chi\left( \overrightarrow{r}\right)
\label{42}
\end{equation}
The system is then decoupled and we get the following second order equation:%
\begin{equation}
\left[ c^{2}p^{2}+2\left( E-\mu c^{2}\right) U\left( \vec{r}\right) -\left(
E^{2}-\mu^{2}c^{4}\right) \right] \chi\left( \overrightarrow{r}\right) =0
\label{43}
\end{equation}

We see that the two equations \ref{39} and \ref{41} are equivalent to the
equations \ref{34}.

We start by studying the spin-symmetry case and then we deduce the solutions
for pseudo-spin symmetry by using the transformation $E\rightarrow -E$ (and
we add $\chi\leftrightarrows\varphi$ for Dirac equation).

\section{Solutions of the Schr\"{o}dinger Type Equation}

\label{2DSR}

\subsection{The Spin Symmetry Case}

The Schr\"{o}dinger type equation for the spin-symmetry case is \ref{39}:%
\begin{equation}
\left[ c^{2}p^{2}+2\left( E+\mu c^{2}\right) U\left( \vec{r}\right) -\left(
E^{2}-\mu^{2}c^{4}\right) \right] \psi\left( \overrightarrow{r}\right) =0
\label{44}
\end{equation}

with the potential energy:%
\begin{equation}
U(r,\theta)=eV(r,\theta)=-\,Z\frac{e^{2}}{r}+e\frac{D_{r}}{r^{2}}+e\,\frac{%
D_{\theta}\cos\theta}{r^{2}}\,  \label{45}
\end{equation}

We use the polar coordinates and the same transformation as before $\psi
(r,\theta )=r^{-1/2}R(r)\Theta (\theta )$ to get two separate equations:
\begin{subequations}
\begin{gather}
\left( \dfrac{\mathrm{d}^{2}}{\mathrm{d}\theta ^{2}}-2\frac{E+\mu c^{2}}{%
\hbar ^{2}c^{2}}eD\cos \theta \right) \Theta (\theta )=E_{\theta }\Theta
(\theta )\,  \label{46a} \\
\left[ \dfrac{\mathrm{d}^{2}}{\mathrm{d}r^{2}}+\left( E_{\theta }-2\frac{%
E+\mu c^{2}}{\hbar ^{2}c^{2}}eD_{r}+\frac{1}{4}\right) \frac{1}{r^{2}}+2%
\frac{E+\mu c^{2}}{\hbar ^{2}c^{2}}Ze^{2}\frac{1}{r}\right] R(r)=-\frac{%
E^{2}-\mu ^{2}c^{4}}{\hbar ^{2}c^{2}}R(r)\,  \label{46b}
\end{gather}%
These equations are equivalent to those of the non-relativistic case \ref%
{11a} and \ref{11b} but we have to consider the shifts \ref{35}. So we use
the same steps to solve \ref{46a} and \ref{46b} and we get two new
eigenvalues relations coming from both angular and radial equations:
\end{subequations}
\begin{subequations}
\begin{gather}
E_{\theta }^{(2m)}=-\frac{1}{4}c_{2m}\left( 4\frac{E_{n,m}+2\mu c^{2}}{\hbar
^{2}c^{2}}eD_{\theta }\right)  \label{47a} \\
E_{\theta }^{(2m)}=2\frac{E_{n,m}+2\mu c^{2}}{\hbar ^{2}c^{2}}eD_{r}-\left(
n-\left\vert m\right\vert +\frac{1}{2}-Z\alpha \frac{E_{n,m}+2\mu c^{2}}{%
\sqrt{\mu ^{2}c^{4}-\left( E_{n,m}+\mu c^{2}\right) ^{2}}}\right) ^{2}
\label{47b}
\end{gather}

We used the non-relativistic energies $E-\mu c^{2}$ and we denoted them $%
E_{n_{r},m}$ or $E_{n,m}$\ where $n=n_{r}+\left\vert m\right\vert $\ ($%
\alpha $ is the fine structure constant).

For the wavefunctions we get the same relation as \ref{22} but with the new
parameters $\beta$ and $\lambda$:
\end{subequations}
\begin{equation}
\beta^{2}=-\frac{E^{2}-\mu^{2}c^{4}}{\hbar^{2}c^{2}}\,\text{\& }\lambda =%
\frac{1}{2}+\sqrt{-E_{\theta}^{(2m)}+2\frac{E+\mu c^{2}}{\hbar^{2}c^{2}}%
eD_{r}}\,  \label{48}
\end{equation}

The same relations are used in the expression of the normalization constant %
\ref{23}.

The non-relativistic limit is obtained by neglecting the term $E_{n,m}$
beside the factor $2\mu c^{2}$ in \ref{47a}, then we replace in \ref{47b}
and we use the Taylor series according to $\alpha^{2}$:%
\begin{equation}
E_{n,m}=-\frac{8\mu c^{2}Z\alpha^{2}}{\left( 2n_{r}+2\sqrt{%
-E_{\theta}^{(2m)}+4\frac{\mu\alpha c}{e\hbar}D_{r}}+1\right) ^{2}}+\frac{%
32\mu c^{2}Z^{2}\alpha^{4}}{\left( 2n_{r}+2\sqrt{-E_{\theta}^{(2m)}+4\frac {%
\mu\alpha c}{e\hbar}D_{r}}+1\right) ^{4}}+O(\alpha^{6})  \label{49}
\end{equation}

This result generalizes the one found in the non-relativistic case \ref{31}
and the ones found for non-relativistic non-pure dipole by putting $D_{r}=0$
\cite{Moumni16}. One can also found the energies of the relativistic Kratzer
potential using $D_{\theta}=0$ or $E_{\theta}^{(2m)}=-m^{2}$ \cite{Saad08}.
The two conditions $D_{r}=0$ and $D_{\theta}=0$\ together give the energies
of the Coulomb potential \cite{Berkdemir07, Arda12}.

We use the Hartree units ($\mu =e=4\pi \epsilon _{0}=\hbar =1$ and $%
c=1/\alpha $) for the numerical computations and we apply for $Z=1$:
\begin{subequations}
\begin{gather}
E_{\theta }^{(2m)}=-\frac{1}{4}c_{2m}\left( 4\left( E_{n,m}\alpha
^{2}+2\right) D_{\theta }\right)  \label{50a} \\
E_{\theta }^{(2m)}=2\left( E_{n,m}\alpha ^{2}+2\right) D_{r}-\left(
n-\left\vert m\right\vert +\frac{1}{2}-Z\alpha \frac{E_{n,m}\alpha ^{2}+2}{%
\sqrt{1-\left( E_{n,m}\alpha ^{2}+1\right) ^{2}}}\right) ^{2}  \label{50b}
\end{gather}

And the non-relativistic limit becomes:
\end{subequations}
\begin{equation}
E_{n,m}=-\frac{2}{\left( n_{r}+\frac{1}{2}+\sqrt{\frac{1}{4}c_{2m}\left(
8D_{\theta }\right) +4D_{r}}\right) ^{2}}+\frac{8\alpha ^{2}}{\left( n_{r}+%
\frac{1}{2}+\sqrt{\frac{1}{4}c_{2m}\left( 8D_{\theta }\right) +4D_{r}}%
\right) ^{4}}+O(\alpha ^{6})  \label{51}
\end{equation}

We see here that \ref{51} differs from \ref{31} by a factor of $2$\ in front
of the dipoles moments $D_{\theta}$\ and $D_{r}$. This factor comes from the
addition of scalar and vector potentials in spin-symmetry case which gives a
Schr\"{o}dinger equation with a potential $2V$ instead of $V$ in ordinary
theory \cite{Alhaidari06, Durmus07}.

We cannot solve the system of equations \ref{50a} and \ref{50b} analytically
because Mathieu characteristics don't have inverse functions. Nevertheless,
this system can be solved using graphical methods by seeking the
intersection points of the graphs representing the two equations.

Equation \ref{50b} shows that $E_{\theta}$ has an inverted and non-symmetric
parabolic shape and the intersection point with the plots representing \ref%
{50a} can not exceed its maximum; This limitation gives the critical dipole
moments for each quantum numbers. Unlike the non-relativistic case where $%
D_{crit}$ depends only of the value of $m$, its values are here weakly
dependent on the other quantum number $n$. This dependence on $n$ comes from
the presence of the energies $E_{n,m}$ with $D$ in the angular eigenvalues %
\ref{50a} and these energies depend on $n$ as can be seen from \ref{50b}.
The weakness of this dependence comes from the presence of the factor $%
\alpha^{2}$ with $E_{n,m}$.

\begin{figure}
\centering
\includegraphics[width=0.5\textwidth]{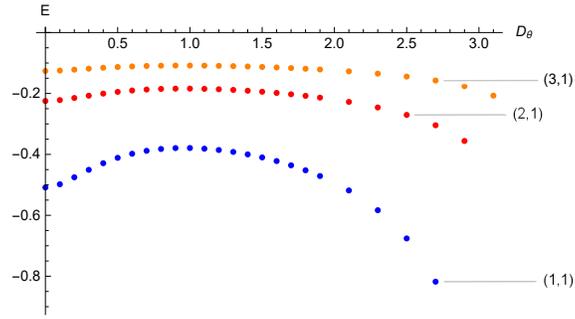}
\caption{Relativistic $E(n,1)$ vs $D_{\protect\theta}$ for $D_{r}=0.3$ and $%
n=1$, $2$ and $3$}
\label{fig7}
\end{figure}

\begin{figure}
\centering
\includegraphics[width=0.5\textwidth]{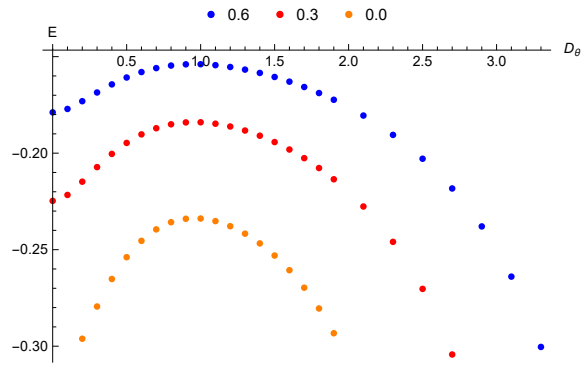}
\caption{$E(2,1)$ vs $D_\protect\theta$ for $D_r=0$ , $0.3$ and $0.6$}
\label{fig8}
\end{figure}

The study of the dependence of the energies according to the values of $%
D_{\theta }$ shows that this moment increases the energies of the system to
a maximum value and then its effect is transformed into a decrease thereof;
This shape follows that of the $c_{2m}$ and it is common to all levels but
decreases with increasing $n$. The effect of $D_{r}$ can be summarized in a
shift of the energies to larger or smaller values depending on its sign
(Figures \ref{fig7} and \ref{fig8}).

\begin{figure}[tbp]
\centering
\includegraphics[width=0.5\textwidth]{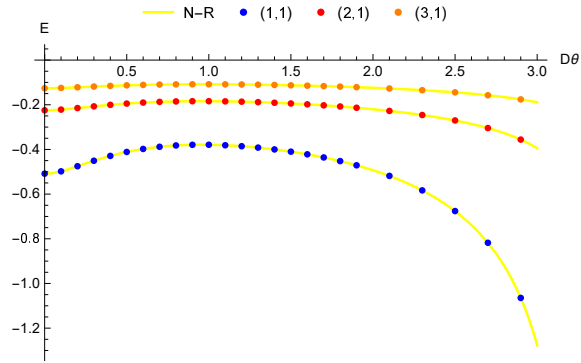}
\caption{Relativistic and Non-Relativistic $E(1,1)$, $E(2,1)$ and $E(3,1)$
vs $D_\protect\theta$ for $D_r=0.3$}
\label{fig9}
\end{figure}

We mention here that the non-relativistic approximation \ref{51} can be used
as a quasi-analytical solution since it gives results in excellent agreement
with those computed numerically (fig\ref{fig9}).

\subsection{The Pseudo-Spin Symmetry Case}

The Schr\"{o}dinger type equation for the pseudo-spin-symmetry case is \ref%
{43}:%
\begin{equation}
\left[ c^{2}p^{2}+2\left( E-Mc^{2}\right) U\left( \vec{r}\right) -\left(
E^{2}-M^{2}c^{4}\right) \right] \psi\left( \overrightarrow{r}\right) =0
\label{52}
\end{equation}

Following the same procedure as that of the spin case, we end up with two
relations that come from the eigenvalues of radial and angular equations. We
find the following relations for the non-relativistic energies of the system
$E_{n,m}=E-\mu c^{2}$ in Hartree units:

\begin{subequations}
\begin{gather}
E_{\theta }^{(2m)}=-\frac{1}{4}c_{2m}\left( 4E_{n,m}\alpha ^{2}D\right)
\label{53a} \\
E_{\theta }^{(2m)}=2E_{n,m}\alpha ^{2}D_{r}-\left( n-\left\vert m\right\vert
+\frac{1}{2}-Z\alpha \frac{E_{n,m}\alpha ^{2}}{\sqrt{1-\left( E_{n,m}\alpha
^{2}+1\right) ^{2}}}\right) ^{2}  \label{53b}
\end{gather}

The main difference between these equations and those of the spin symmetry
case (\ref{50a}, \ref{50b}) is the absence of factor $2$ in front of the $%
\alpha^{2}$ term. This means that the graphs representing the radial
solution \ref{53b} are almost linear and that the parameter $p$ inside the
Mathieu characteristics \ref{53a} is very small. Our calculations show that
we have to consider very large radial moments ($D_{r}>100a.u.$) to find
solutions higher than $-200a.u.$. These results are outside the regions of
interest for the energies of the atomic systems and support those of works
that consider only the case of spin symmetry in their studies \cite{Saad08,
Durmus07, Ikhdair08}.

\section{Discussion}

\label{Concl}In this work, we studied the potential $V\left( r,\theta \right)
=Qr^{-1}+D_{r}r^{-2}+D_{\theta }\cos (\theta )r^{-2}$ for $2D$ quantum
systems in both non-relativistic and relativistic cases. We solved the Schr%
\"{o}dinger equation analytically and studied the relativistic spectrum for
Klein-Gordon and Dirac equations in both spin and pseudo-spin symmetry cases.

In the non-relativistic case, the spectrum shows that the energies follow
mainly the behavior of Mathieu's characteristic parameters and thus the
angular moment $D_{\theta }$, whereas the effect of the radial moment $D_{r}$
is merely a shift in these energies to larger or smaller values according to
its sign. We have showed also that there is an essential condition for bound
states to exist, which is: $c_{2m}\left( 4D_{\theta }\right) +8D_{r}>0$.
This condition imposes a critical value for the angular moment $D_{\theta }$%
, depending on the value of $m$, otherwise the corresponding bound state
disappears. These critical values of $D_{\theta }$ depend also on the value
of $D_{r}$ and the negative value of this moment which makes $c_{2m}\left(
4D_{\theta }\right) +8D_{r}=0$ is also a critical value for the radial
moment. So we see that by increasing, the radial dipole displaces the
energies towards the larger values while widening the region of the possible
values of the angular moment.

In the relativistic cases the eigenfunctions are determined analytically but
the energies can only be calculated using graphical methods. Only the spin
symmetry has given results corresponding to atomic systems. The behavior of
the energies is the same as that of the Schr\"{o}dinger spectrum but it is
shifted because the Schr\"{o}dinger type equation of the relativistic
systems has $2V$ as a potential instead of the potential $V$ in the ordinary
Schr\"{o}dinger equation. We also note that the critical values of the
dipole moments $D_{r}$ and $D_{\theta }$ depend on the two quantum numbers $n
$ and $m$ in the relativistic case instead of just $m$ in the case of
non-relativistic systems.

We have found that the angular term removes the degeneracy found in the $%
exp(im\theta )$ part of the solutions for central potentials. This is
equivalent to the effect of a constant magnetic field in $3D$ systems, where
its action removes the degeneracy of the $exp(im\varphi )$ solutions too. In
both cases, the privileged direction of the interaction (dipole axis in $2D$
and field direction in $3D$) removes the degeneracy that existed due to the
isotropy of the action before.

\section*{Acknowledgement}
The authors would like to thank Pr Y. Delenda for his help in writing this work and also Dr. Y. Gindikin for his comment which has enriched it.

\let\chapter\undefined%
\appendix{}
\end{subequations}


\begin{thebibliography}{99}
\bibitem{Khare94} Khare A. and Bhaduri R.K.; \textit{Supersymmetry, Shape
Invariance and Exactly Solvable Noncentral Potentials}, Am. J. Phys. \textbf{%
62}, 1008-1014 (1994)

\bibitem{Arda12} Arda A. and Sever R.; \textit{Non-central potentials, exact
solutions and Laplace transform approach}, J. Math. Chem. \textbf{50},
1484--1494 (2012)

\bibitem{AlHaidari18} Alhaidari A.D. and Bahlouli H.; \textit{Electric
dipole and quadrupole contributions to valence electron binding in a
charge-screening environment}, Eur. Phys. J. D \textbf{73}, 70 (2019)

\bibitem{Kumari18} Kumari N., Yadav R.K., Khare A. and Mandal B.P.; \textit{%
A class of exactly solvable rationally extended non-central potentials in
two and three dimensions}, J. Math. Phys. \textbf{59}, 062103 (2018)

\bibitem{Hautot73} Hautot A.; \textit{Exact motion in noncentral electric
fields}, J. Math. Phys. \textbf{14}, 1320 (1973)

\bibitem{Makarov67} Makarov A.A., Smorodinsky J.A., Valiev Kh. and
Winternitz P.; \textit{A systematic search for nonrelativistic systems with
dynamical symmetries}, Nuovo Cimento A \textbf{52}, 1061 (1967)

\bibitem{Hartmann72} Hartmann H.; \textit{Die Bewegung eines K\"{o}rpers in
einem ringf\"{o}rmigen Potentialfeld}, Theor. Chim. Acta \textbf{24}, 201
(1972)

\bibitem{Gharbi13} Gharbi A. and Bouda A.; \textit{Energy spectra of
Hartmann and ring-shaped oscillator potentials using the quantum
Hamilton--Jacobi formalism}, Phys. Scr. \textbf{88}, 045007 (2013)

\bibitem{Bharali13} Bharali A.; \textit{Systematic search of exactly
solvable ring-shaped potential using the transformation method, }Phys. Scr.
\textbf{88}, 035009 (2013)

\bibitem{Sun14} Dong-Sheng Sun, Yuan You, Fa-Lin Lu, Chang-Yuan Chen and
Shi-Hai Dong; \textit{The quantum characteristics of a class of complicated
double ring-shaped non-central potential}, Phys. Scr. \textbf{89}, 045002
(2014)

\bibitem{Gribakin15} Gribakin G.F. and Swann A.R., \textit{Effect of dipole
polarizability on positron binding by strongly polar molecules}, J. Phys. B
\textbf{48}, 215101 (2015)

\bibitem{Fermi47} Fermi E. and Teller E., \textit{The Capture of Negative
Mesotrons in Matter}, Phys. Rev. \textbf{72,} 399 (1947)

\bibitem{Turner77} Turner J.E.; \textit{Minimum dipole moment required to
bind an electron-molecular theorists rediscover phenomenon mentioned in
Fermi-Teller paper twenty years earlier}, Am. J. Phys. \textbf{45}, 758
(1977)

\bibitem{Fox66} Fox K. and Turner J.E.; \textit{Variational Calculation for
Bound States in an Electric-Dipole Field}, J. Chem. Phys. \textbf{45}, 1142
(1966)

\bibitem{Crawford67} Crawford O. and Dalgarno A.; \textit{Bound states of an
electron in a dipole field}, Chem. Phys. Lett. \textbf{1}, 23 (1967)

\bibitem{Gutsev95} Gutsev G.L. and Adamowicz L.; \textit{Electronic and
geometrical structure of dipole-bound anions formed by polar molecules}, J.
Phys. Chem. \textbf{99}, 13412--13421 (1995)

\bibitem{Jordan03} Jordan K.D. and Wang F.; \textit{Theory of Dipole-Bound
Anions}, Annu. Rev. Phys. Chem. \textbf{54}, 367-396 (2003)

\bibitem{Svozil04} Svozil D., Jungwirth P. and Havlas Z.; \textit{Electron
Binding to Nucleic Acid Bases. Experimental and Theoretical Studies. A Review%
}, Collect. Czech. Chem. Commun. \textbf{69}, 1395-1428 (2004)

\bibitem{Simons08} Simons J.; \textit{Molecular Anions}, J. Phys. Chem. A
\textbf{112}, 6401--6511 (2008)

\bibitem{Connoly07} Connolly K. and Griffiths D.J., \textit{Critical dipoles
in one, two, and three dimensions}, Am. J. Phys. 75 6 (2007)

\bibitem{Glasser07} Glasser M.L. and Nieto L.M.; \textit{Electron capture by
an electric dipole in two dimensions}, Phys. Rev. A \textbf{75}, 062109
(2007)

\bibitem{AlHaidari08} AlHaidari A.D.; \textit{Analytic Solution of the Schr%
\"{o}dinger Equation for an Electron in the Field of a Molecule with an
Electric Dipole Moment}, Ann. Phys. \textbf{323} 1709 (2008)

\bibitem{Moumni16} Moumni M. and Falek M.; \textit{Schr\"{o}dinger Equation
for Non-Pure Dipole Potential in 2D Systems}, J. Math. Phys. \textbf{57},
072104 (2016)

\bibitem{Gorlitz01} G\"{o}rlitz A;. et al.; \textit{Realization of
Bose-Einstein Condensates in Lower Dimensions}, Phys. Rev. Lett. \textbf{87}%
, 130402 (2001)

\bibitem{Martiyanov10} Martiyanov K., Makhalov V. and Turlapov A.; \textit{%
Observation of a Two-Dimensional Fermi Gas of Atoms}, Phys. Rev. Lett.
\textbf{105,} 030404 (2010)

\bibitem{Safonov98} Safonov, A.I., Vasilyev S.A., Yasnikov I.S., Lukashevich
I.I. and Jaakkola S.; \textit{Observation of Quasicondensate in
Two-Dimensional Atomic Hydrogen}, Phys. Rev. Lett. \textbf{81}, 4545-4548
(1998)

\bibitem{Zhou90} Zhou J.L. and Xiong J.J.; \textit{Hydrogen molecular ions
in two dimensions}, Phys. Rev. B \textbf{41}, 12274-12277 (1990)

\bibitem{Vasilyev02} Vasilyev S., Jarvinen J., Safonov
A.I., Kharitonov A.A., Lukashevich I.I. and Jaakkola S.; \textit{%
Electron-Spin-Resonance Instability in Two-Dimensional Atomic Hydrogen Gas},
Phys. Rev. Lett. \textbf{89, }153002 (2002)

\bibitem{Gadella11} Gadella M., Negro J., Nieto L.M. and Pronko G.P.;
\textit{Two Charged Particles in the Plane Under a Constant Perpendicular
Magnetic Field}, Int. J. Theor. Phys. \textbf{50}, 2019-2028 (2011)

\bibitem{Martino14} De Martino A., Kl\"{o}pfer D., Matrasulov D.U. and Egger
R.; \textit{Electric-Dipole-Induced Universality for Dirac Fermions in
Graphene}, Phys. Rev. Lett. \textbf{112}, 186603 (2014)

\bibitem{Klopfer14} Kl\"{o}pfer D., De Martino A., Matrasulov D.U. and Egger
R.; \textit{Scattering theory and ground-state energy of Dirac fermions in
graphene with two Coulomb impurities}, Eur. Phys. J. B \textbf{87}, 187
(2014)

\bibitem{Gindikin18} Gindikin Y. and Sablikov V. A.; \textit{%
Spin-orbit-driven electron pairing in two dimensions}, Phys. Rev. B \textbf{%
98}, 115137 (2018)

\bibitem{Kratzer20} Kratzer A.; \textit{Die ultraroten Rotationsspektren der
Halogenwasserstoffe}, Z. Phys. \textbf{3}, 289-307 (1920)

\bibitem{Kratzer26} Kratzer A.; \textit{Die Gesetzm\"{a}ssigkeiten in den
Bandspektren}, Enc. d. Math. Wiss. \textbf{3}, 821-859 (1926)

\bibitem{Fortunato03} Fortunato L. and Vitturi A.; \textit{Analytically
solvable potentials for gamma unstable nuclei}, J.Phys. G \textbf{29,}
1341-1350 (2003)

\bibitem{Hagigeorgiou06} Hajigeorgiou P.G.; \textit{Exact analytical
expressions for diatomic rotational and centrifugal distortion constants for
a Kratzer Fues oscillator}, J. Molec. Spect. \textbf{235}, 111-116 (2006)

\bibitem{Berkdemir06} Berkdemir C., Berkdemir A. and Han J.; \textit{Bound
state solutions of the Schr\"{o}dinger equation for modified Kratzer's
molecular potential}, Chem. Phys. Lett. \textbf{417}, 326-329 (2006)

\bibitem{Hooydonk09} Van Hooydonk G.; \textit{Ionic Kratzer bond theory and
vibrational levels for achiral covalent bond HH}, Z. Naturforsch. A \textbf{%
64}, 801 (2009)

\bibitem{Batra18} Batra K. and Prasad V.; \textit{Spherical quantum dot in
Kratzer confining potential: study of linear and nonlinear optical
absorption coefficients and refractive index change}, Eur. Phys. J. B 91,
298 (2018)

\bibitem{Cheng07} Cheng Y.F. and Dai T.Q.; \textit{Exact solution of the Schr%
\"{o}dinger equation for the modified Kratzer potential plus a ring-shaped
potential by the Nikiforov--Uvarov method, }Phys. Scr. \textbf{75, }274--277
(2007)

\bibitem{Babaei11} Babaei-Brojeny A.A. and Mokari M.; \textit{An analysis of
the applications of the modified Kratzer potential, }Phys. Scr. \textbf{84 }%
045003 (2011)

\bibitem{Molas19} Molas M.R., Slobodeniuk A.O., Nogajewski K., Bartos M.,
Bala L., Babinski A., Watanabe K., Taniguchi T., Faugeras C. and
Potemski M.; \textit{Energy spectrum of two-dimensional excitons in a
non-uniform dielectric medium}, arXiv:1902.03962

\bibitem{Sivoukhine86} Sivoukhine D.; \textit{Cours de physique g\'{e}n\'{e}%
rale - Tome 5. Physique atomique et nucl\'{e}aire - Premi\`{e}re partie};
French Traduction Editions MIR pp 223 (1986) ASIN: B00AW98XK8

\bibitem{Molski92} Molski M. and Konarski J.; \textit{Modified kratzer-fues
formula for rotation-vibration energy of diatomic molecules}, Acta. Phys.
Pol. A \textbf{82}, 927-936 (1992)

\bibitem{Durmus11} Durmus A.; \textit{Non-relativistic treatment of diatomic
molecules interacting with a generalized Kratzer potential in hyperspherical
coordinates, }J. Phys. A: Math. Theor. \textbf{44,} 155205 (2011)

\bibitem{Bao19} Bao J. and ShizgalB.D.; \textit{Pseudospectral method of
solution of the Schr\"{o}dinger equation for the Kratzer and pseudoharmonic
potentials with nonclassical polynomials and applications to realistic
diatom potentials}, Comp. Theor. Chem. \textbf{1149}, 49-56 (2019)

\bibitem{Mathieu} Mathieu E., \textit{M\'{e}moire sur le mouvement
vibratoire d'une membrane de forme elliptique}, J. Math. Pures. Appl.
\textbf{13,} 137 (1868)

\bibitem{Abram72} Abramowitz M. and Stegun I.A., \textit{Handbook of
Mathematical Functions}, Dover Publ., New York, (1972)

\bibitem{Floquet} Floquet G., \textit{Sur les \'{e}quations diff\'{e}%
rentielles lin\'{e}aires \`{a} coefficients p\'{e}riodiques}, Annales de l'%
\'{E}cole Normale Sup\'{e}rieure \textbf{12} 47 (1883)

\bibitem{Bloch28} Bloch F., \textit{\"{U}ber die Quantenmechanik der
Elektronen in Kristallgittern}, Z. Physik \textbf{52} 555 (1928)

\bibitem{NIST} https://dlmf.nist.gov/28.6 (2018-12-15)

\bibitem{Saad03} Saad N. and Hall R.L.; \textit{Integrals containing
confluent hypergeometric functions with applications to perturbed singular
potentials}, J. Phys. A: Math. Gen. \textbf{36,} 7771--7788 (2003)

\bibitem{Oyewumi05} Oyewumi K.J.; \textit{Analytical Solutions of the
Kratzer-Fues Potential in an Arbitrary Number of Dimensions}, Found. Phys.
Lett. \textbf{18}, 75 (2005)

\bibitem{Agboola11} Agboola A.; \textit{Complete Analytic Solutions of the
Mie-type Potentials in N-Dimensions}, Acta. Phys. Polonica A \textbf{120},
371 (2011)

\bibitem{Zaslow67} Zaslow B. and Zandler M.E., \textit{Two-Dimensional
Analog to the Hydrogen Atom}, Am. J. Phys. \textbf{35,} 1118 (1967)

\bibitem{Parfitt02} Parfitt D.G.W. and Portnoi M.E., \textit{The
two-dimensional hydrogen atom revisited}, J. Math. Phys. \textbf{43}, 4681
(2002)

\bibitem{Hecht69} Hecht K.T. and Adler A.; \textit{Generalized seniority for
favored }$J\neq0$\textit{\ pairs in mixed configurations}, Nucl. Phys. A.
137 129-143 (1969)

\bibitem{Arima69} Arima A., Harvey M. and Shimizu K.; \textit{Pseudo LS
coupling and pseudo SU(3) couling shemes}, Phys. Lett. B. 30 517-522 (1969)

\bibitem{Ginocchio97} Ginocchio J.N.; \textit{Pseudospin as a Relativistic
Symmetry}, Phys. Rev. Lett. 78, 436 (1997)

\bibitem{Zhou03} Zhou S.G., Meng J. and Ring P.; \textit{Spin Symmetry in
the Antinucleon Spectrum}, Phys. Rev. Lett. 91, 262501 (2003)

\bibitem{Liang15} Liang H., Meng J. and Zhou S.G.; \textit{Hidden pseudospin
and spin symmetries and their origins in atomic nuclei}, Phys. Rep. 570 1-84
(2015)

\bibitem{Hall10} Hall R.L. and Ye\c{s}ilta\c{s} \"{O}.; \textit{Comparison
theorems for the Dirac equation with spin-symmetric and
pseudo-spin-symmetric interactions}, J. Phys. A Math. Theor. \textbf{43,}
195303 (2010)

\bibitem{Berkdemir08} Berkdemir C. and Sever R.; \textit{Pseudospin symmetry
solution of the Dirac equation with an angle-dependent potential}, J. Phys.
A Math. Theor.\textbf{41,} 045302\textbf{\ }(2008)

\bibitem{Pauli27} Pauli W. Jr.; \textit{Zur Quantenmechanik des magnetischen
Elektrons}, Z. Phys. \textbf{43}, 601 (1927)

\bibitem{Snyder40} Snyder H. and WeinbergJ., \textit{Stationary States of
Scalar and Vector Fields}, Phys. Rev. 57, 307 (1940)

\bibitem{Shiff40} Snyder H. and WeinbergJ., \textit{On The Existence of
Stationary States of the Mesotron Field}, Phys. Rev. 57, 315 (1940)

\bibitem{Schulze17} Schulze-Halberg A. and Roy P., \textit{Bound states of
the two-dimensional Dirac equation for an energy-dependent hyperbolic Scarf
potential} , J. Math. Phys. \textbf{58}, 113507 (2017)

\bibitem{Schulze18} Schulze-Halberg A. and Ye\c{s}ilta\c{s} \"{O}.; \textit{%
Generalized Schr\"{o}dinger equations with energy-dependent potentials:
Formalism and applications}, J.Math.Phys. \textbf{59}, 113503 (2018)

\bibitem{Benzair18} Benzair, H. Merad and M. Boudjedaa, T.; \textit{Electron
propagator with vector and scalar energy-dependent potentials in }(2+1)%
\textit{-dimensional space-time}, Int. J. Mod. Phys. A \textbf{33} 1850186
(2018)

\bibitem{Grad07} Gradshteyn I.S. and Ryzhik I.M., \textit{Table of
Integrals, Series, and Products}, Alan Jeffrey and Daniel Zwillinger (eds.)
Elsevier, London (2007)

\bibitem{Saad08} Saad N., Hall R.L. and Ciftci H.; \textit{The Klein-Gordon
equation with the Kratzer potential in d dimensions}, Cent. Eur. J. Phys.
\textbf{6}, 717-729 (2008)

\bibitem{Berkdemir07} Berkdemir C., \textit{Relativistic treatment of a
spin-zero particle subject to a Kratzer-type potential}, Am. J. Phys. 75, 81
(2007)

\bibitem{Alhaidari06} Alhaidari A.D., Bahlouli H. and Al-Hasan A.; \textit{%
Dirac and Klein Gordon equations with equal scalar and vector potentials},
Phys. Lett. A \textbf{349}, 87-97 (2006)

\bibitem{Durmus07} Durmus A. and Yasuk F.; \textit{Relativistic and
nonrelativistic solutions for diatomic molecules in the presence of double
ring-shaped Kratzer potential}, J. Chem. Phys. \textbf{126}, 074108 (2007)

\bibitem{Ikhdair08} Ikhdair S.M. and Sever R.; \textit{Relativistic solution
in D-dimensions to a spin-zero particle for equal scalar and vector
ring-shaped potential, }Cent. Eur. J. Phys. \textbf{6}, 141-152 (2008)
\end{thebibliography}
\end{document}